\documentclass[aps,prb,amsmath,amssymb]{revtex4-2}
\usepackage{graphicx}
\usepackage{amsmath}
\usepackage{braket}
\usepackage{amssymb}
\usepackage{multirow}
\usepackage{bbm}
\usepackage{bm}
\usepackage{mathtools}
\usepackage{hyperref}
\usepackage{natbib}
\usepackage{mathtools}
\usepackage{enumerate}
\usepackage[normalem]{ulem}

\usepackage[x11names]{xcolor}

\begin{document}

\title{Electron Density and Compressibility  in the Kitaev Model with a Spatially Modulated Phase in the Superconducting Pairing}

 	\author{Fabian G. Medina Cuy}	\email{fabian.medina@polito.it}
	\author{Fabrizio Dolcini}
	\affiliation{Dipartimento di Scienza Applicata e Tecnologia, Politecnico di Torino, corso Duca degli Abruzzi 24, 10129 Torino (Italy)}
	\date{\today}

\begin{abstract}
	A current flowing through a one-dimensional Kitaev chain induces a spatial modulation in its superconducting pairing, characterized by a wave vector $Q$, which is known to induce two types of topological phase transitions: one is the customary band topology transition between gapped phases, while the other is a Lifshitz transition related to the  Fermi surface topology  and leading to a gapless superconducting phase. We investigate the behavior of the electron density $\rho$ and the compressibility $\kappa$ across the two types of transitions, as a function of the model parameters.  We find that the  behavior of $\rho$ as a function of $Q$ and chemical potential $\mu$ enables one to infer the ground state phase diagram. Moreover, the analysis of the compressibility $\kappa$ as a function of $\mu$ enables one to distinguish the two transitions: While $\kappa$ exhibits a symmetric divergence across the band topology transition, it displays an asymmetric jump across the Lifshitz transition.

	\end{abstract}

	\maketitle

\setcounter{section}{0} 

\section{Introduction}
Topological superconductors (TSs) are considered to be extremely promising materials for frontier research in quantum science and technology.  On the one hand, they exhibit Majorana zero modes (MZMs),  exotic quasiparticles with nonlocal correlations and braiding properties that are suitable for fault-tolerant quantum computation. On the other hand, TSs are also characterized by dissipationless transport, which is an ideal feature for the development of energetically sustainable nanoelectronics~\cite{kitaev2003,alicea2012,aguado2017,zhang2018,dassarma2023,aghaee2023}. Various platforms have been proposed for the experimental realization of  1D TSs, including proximized spin-orbit nanowires \cite{lutchyn2010, oreg2010}, quantum spin Hall edges \cite{fu-kane2008, fu-kane2009}, and ferromagnetic atom chains \cite{choy2011, nadjperge2013, glazman2013, franz2013, kotetes2014}. On the theory side,  the essential properties of a $p$-wave TS are considered to be well captured by the  Kitaev chain  model, which exhibits two topologically distinct gapped phases,  with MZMs appearing  at the chain edges when the system is in the   topologically non-trivial phase. Edge correlations  therefore represent fingerprints of the topological phase transitions~\cite{zhou_PRL_2017,chen_PRB_2017,zhou_scirep_2018,murakami_PRB_2021}. Generalizations of the Kitaev chain model including long-range hopping and superconducting terms have also been investigated, and can lead to algebraic decays of the correlation function in gapped phases \cite{ercolessi_PRL_2014, ercolessi_NJP_2016, dellanna_PRB_2020, dellanna_PRB_2022}.\\

Experiments on superconductor/semiconductor nanowires have   focused on the search for zero-bias conductance peaks as the smoking gun evidence of MZMs, requiring an electrical current flow through the system. This has prompted researchers to investigate the effects of a spatially modulated phase in the superconducting pairing, where the  modulation wavevector  $Q$  is proportional to the net momentum carried by a Cooper pair, which is non-vanishing in the presence of a current flow \cite{takasan2022, kotetes2022, ma2023, maiellaro2023}. Notably, a recent work showed that in the physically realistic regime where the superconducting pairing strength $\Delta_0$ is smaller than the bandwidth energy parameter $w$ ($\Delta_{0} < w$), the phase modulation induces two types of phase transitions in the system: the first one is a transition in the band topology of the model, while the second one is a Lifshitz transition~\cite{volovik2017,volovik2018}  from a gapped to a gapless superconducting phase \cite{yerin2022lifshitz, yerin2022topological}. Moreover, by treating $Q$ as the wavector  of a synthetic dimension, a mapping has been established between the current carrying state of the 1D Kitaev model and the ground state of a 2D Weyl semimetal, where the gapless superconducting phase of the former corresponds to a type-II Weyl semimetal in the latter\cite{FFF2024}. Another study   analyzed the long-distance behavior of the correlation functions, showing that the system exhibits different types of exponential decays in the gapped phases and finding the period of the oscillatory algebraic decay in the gapless phases~\cite{FF2024}. Furthermore, a connection was established between the gapless superconducting phase of the Kitaev chain and the chiral phase of spin models with Dzyaloshinskii-Moriya interaction~\cite{FF2024}. \\

Motivated by such promising results, in this work  we focus on  two experimentally accessible  quantities, namely  the electron density and the compressibility, and analyze their behavior across the two types of topological transitions.
The article is organized as follows. Section \ref{sec2} describes the model and briefly summarizes the known properties of  the ground state phase diagram as a function of $Q$ and the chemical potential, recalling the parameter ranges for the gapped and gapless phases of the model. In Section \ref{sec3} we present  our results about the behavior of the density and compressibility across    the boundaries of the two topological phase transitions. Section \ref{sec4} is devoted to the discussion of the results and to   our conclusions.

\section{Materials and Methods}
\label{sec2}
In this section, we briefly describe the model and summarize the main methodological aspects that are needed to investigate the behavior of the density and compressibility. More technical details can be found in Refs.\cite{FFF2024,FF2024}. 
\subsection{Model and Ground State}
 \label{sec2a} 
The Kitaev chain model with a phase modulation of the superconducting $p$-wave pairing is described by the following second quantized Hamiltonian
\begin{equation}
	\mathcal{H} = \sum_{j}\left[ w\left(c^{\dagger}_{j}c_{j+1} + c^{\dagger}_{j+1}c_{j}\right)- \mu \left(c^{\dagger}_{j}c_{j} - \frac{1}{2}\right)  + \Delta_{0}\left( e^{-iQ\left(2j + 1\right)}c^{\dagger}_{j}c^{\dagger}_{j+1} + e^{iQ\left(2j + 1\right)} c_{j+1}c_{j}   \right) \right],
	\label{Ham-Kitaev}
\end{equation} 
where  $c^\dagger_{j}$ and $c^{}_{j}$ are electron creation and annihilation operators at the lattice site $j$. Here, $w>0$ is the inter-site  hopping energy, $\mu$  the chemical potential, $\Delta_{0}>0$  the strength of the superconducting pairing, while $Q$ is the wavevector characterizing its spatial modulation and describing   a current flow along the chain. We assume to deal with an infinitely long chain, where the number of sites is $N_{s} \gg 1$. By  Fourier transform, we can rewrite the Hamiltonian (\ref{Ham-Kitaev}) as
\begin{equation}
	\mathcal{H} = \frac{1}{2}\sum_{k} \Psi^{\dagger}_{k;Q} H(k,Q) \Psi_{k;Q},
	\label{kitaev_model_k}
\end{equation}
where  $\Psi^{\dagger}_{k;Q} = (c^{\dagger}_{k-Q} \,,\, c_{-k-Q})$ is the Nambu spinor, while the $2 \times 2$ matrix
\begin{equation}
	H(k,Q) = h_{0}(k,Q)\sigma_{0} + \vec{h}(k,Q)\cdot\vec{\sigma}	\label{Bloch_Hamiltonian}
\end{equation}
is the  Bogoliubov-de Gennes Hamiltonian, 
where $\sigma_{0}$ is the  identity matrix, $\vec{\sigma}$ is the vector $(\sigma_{1},\, \sigma_{2},\, \sigma_{3})$ of Pauli matrices and
\begin{align}
	h_{0}(k,Q) =& 2w\sin\left(Q\right)\sin\left(k\right), \\
	\vec{h}(k,Q) =& \left(0 \, , \,\text{Im}\left\{\Delta(k)\right\} \, , \,\xi(k,Q)\right) 
\end{align}
with
\begin{align}
	\Delta(k) =& 2\Delta_{0} i \sin(k) \label{Delta(k)}\\
	\xi(k,Q) =& 2w\cos(Q)\cos(k) -\mu \quad.
\end{align}
The single particle spectrum consists of an upper band $E_{+}$ and a lower band $E_{-}$,  given by  
\begin{equation}
	E_{\pm}(k,Q) = h_{0}(k,Q) \pm \sqrt{ \xi^{2}(k,Q) + \left|\Delta(k)\right|^{2}  } \quad.
	\label{Eigenvalues}
\end{equation}
Here, $\Delta(k)$ is odd under $k \rightarrow -k$, as clearly shown by Eq.(\ref{Delta(k)}), and has a $p$-wave character. Thus, differently from $s$-wave superconductors,  $\Delta_0$ cannot be identified with the minimal spectral gap, which in general depends on $\Delta_0$, $\mu$ and $Q$. This can be illustrated in the simple case of the customary Kitaev model ($Q=0$). A few algebra steps show that, in the regime $\Delta_0>w$, the minimal gap $\Delta_g$ of the spectrum (\ref{Eigenvalues}) occurs at $k=0$ for $\mu>0$ or at $k=\pm \pi$ for $\mu<0$, and is given by $\Delta_g=2 |2w-|\mu||$, regardless of the specific value of $\Delta_0$. In contrast,  in the regime $\Delta_0<w$, the minimal gap occurs at a $\pm k^*$ (with $0<|k^*|<\pi$) and depends also on the specific value of $\Delta_0$. In particular, for $\mu=0$, one has $\Delta_g=4 \Delta_0$. 

As discussed in Refs.\cite{FFF2024,FF2024}, for fixed values of $\Delta_0$ and $\mu$, depending on the value of the spatial modulation $Q$, the spectrum~(\ref{Eigenvalues}) can be either gapped, or characterized by a direct gap closing occurring at $k=0,\pi$, or even exhibit an indirect gap closing, if $Q$ is sufficiently large. In the latter case, the occupancy of the bands changes, modifying the physical nature of the ground state from a gapped to a gapless superconductor.
Thus,  the  ground state is in general characterized by three sectors of the Brillouin zone: the Cooper pair sector $S_p$, the unpaired electron sector $S_e$, and the unpaired hole sector  $S_{h}$. These three sectors form the entire Brillouin zone as ${\rm BZ} \equiv S_{p}\cup S_{e} \cup S_{h}$. We will also refer to the  unpaired sector   as the region  $S_{u} = S_{e} \cup S_{h}$. The ground state   takes the general expression
\begin{equation}
	\ket{G(Q)} = \prod_{\substack{0< k < \pi \\ k \in S_{p}}}\left(u_{Q}(k) + v^{*}_{Q}(k)c^{\dagger}_{-k-Q}c^{\dagger}_{k-Q}\right) \prod_{k \in S_{e}} c^{\dagger}_{k-Q} \ket{0} \quad,
	\label{ground_state}
\end{equation}
where 
\begin{equation}
	u_{Q}(k) = \sqrt{\frac{1}{2}\left(1 + \frac{\xi(k,Q)}{h(k,Q)}\right)}, \quad v_{Q}(k) = -i\text{sgn}\left(\sin(k)\right)\sqrt{ \frac{1}{2}\left( 1 - \frac{\xi(k,Q)}{h(k,Q)}  \right)  } 
\end{equation} 
are the weights of the   eigenstates  $(u_{Q}(k), \,\, v_{Q}(k))^{T}$ and $(-v^{*}_{Q}(k),\,\, u_{Q}(k))^{T}$ of Eq.(\ref{Bloch_Hamiltonian}).  \\
Before concluding  this subsection, we would like to highlight a few differences 
between our model Eq.(\ref{Ham-Kitaev}) and the model for a FFLO state~\cite{FF,LO}. Although in both cases the superconducting pairing is charactarized by a spatial modulation, in the FFLO case its wavevector $Q$ is related to an exchange field, typically due to ferromagnetic impurities and acting on the electron spin, whereas here we are considering a {\it spinless} model, and $Q$ originates from a current flow. Moreover,  the FFLO is an equilibrium state, whereas Eq.(\ref{ground_state}) describes a current-carrying (stationary) out of equilibrium state. As a consequence, the spatially modulated pairing is real in the FFLO case~\cite{machida_1984,buzdin_1987,matsuda_2007}, while in Eq.(\ref{Ham-Kitaev})  it exhibits a complex phase, yielding a current.

\subsection{Gapped and Gapless phases in the Kitaev model}
\label{sec2b}
One can identify two distinct regimes, depending on the value of the pairing strength $\Delta_0$. For $\Delta_0>w$ only  gapped phases essentially exist. This is illustrated in Figure \ref{fig_1} (a), where the trivial (topological) gapped phase is denoted in grey (pink) color, as a function of $Q$ and~$\mu$. The two gapped phases are separated by two {\it separatrix curves} (black lines), along which the spectral gap closes {\it directly}. 
In contrast, for $\Delta_0<w$, the spectral gap closes {\it indirectly} and the gapless phase emerges in the $Q$-$\mu$ phase diagram, as highlighted in green in \ref{fig_1} (b). 

\subsubsection{Gapped Phases}
\label{subsec-2.2.1} 
Gapped phases   are identified by the condition that $E_{+}(k,Q) > 0$ and $E_{-}(k,Q) < 0$ for all $k \in {\rm BZ}$. In this case, it can be shown~\cite{FF2024} that the paired sector covers the entire Brillouin zone, while the unpaired sectors are empty
\begin{equation}
\begin{array}{l}
S_p  \equiv {\rm BZ} \leftrightarrow k \in [-\pi,\pi]\\ \\
S_e = S_h = \emptyset
\end{array} \quad. \label{sectors-gapped}
\end{equation}
Then, the general expression  (\ref{ground_state}) of the ground state reduces to the standard form consisting of Cooper pairs only.
The system is in a gapped phase if  one of the three following conditions is met

\begin{equation}
	\begin{array}{ccccl}
		(1) && \left|\mu\right| > 2w &\text{and}& \forall \Delta_{0} > 0 \,\, \text{and} \,\, \forall Q, \\
		\\
		(2) && \left|\mu\right| <2w &\text{and}& \sqrt{w^{2} - \mu^2/4} <\Delta_{0} \,\, \text{and} \,\, \left|\cos(Q)\right| \neq \left|\mu\right|/2w,\\
		\\
		(3) && \left|\mu\right| <2w &\text{and}& w\left|\sin(Q)\right| <\Delta_{0}< \sqrt{w^{2} - \mu^2/4} \quad.
	\end{array}
    \label{gapped_phases}
\end{equation}

As one can notice from the phase diagram in \ref{fig_1}, the condition $(1)$ in Eq.(\ref{gapped_phases}) corresponds to a trivial phase of the system, conditions $(2)$ and $(3)$ can be either topological or trivial. Note, that the second condition in (\ref{gapped_phases}) excludes the separatrix curves, given by  
\begin{equation}
 \mu^\pm_{c}(Q)=\pm 2w \cos{Q} 
\label{separatrix}
\end{equation}
identified by the solid and dashed black lines of the phase diagrams in Figure \ref{fig_1}.

\subsubsection{Gapless Phases}
\label{subsec-2.2.2}
\begin{figure}
    \centering
	\includegraphics[scale = 0.5]{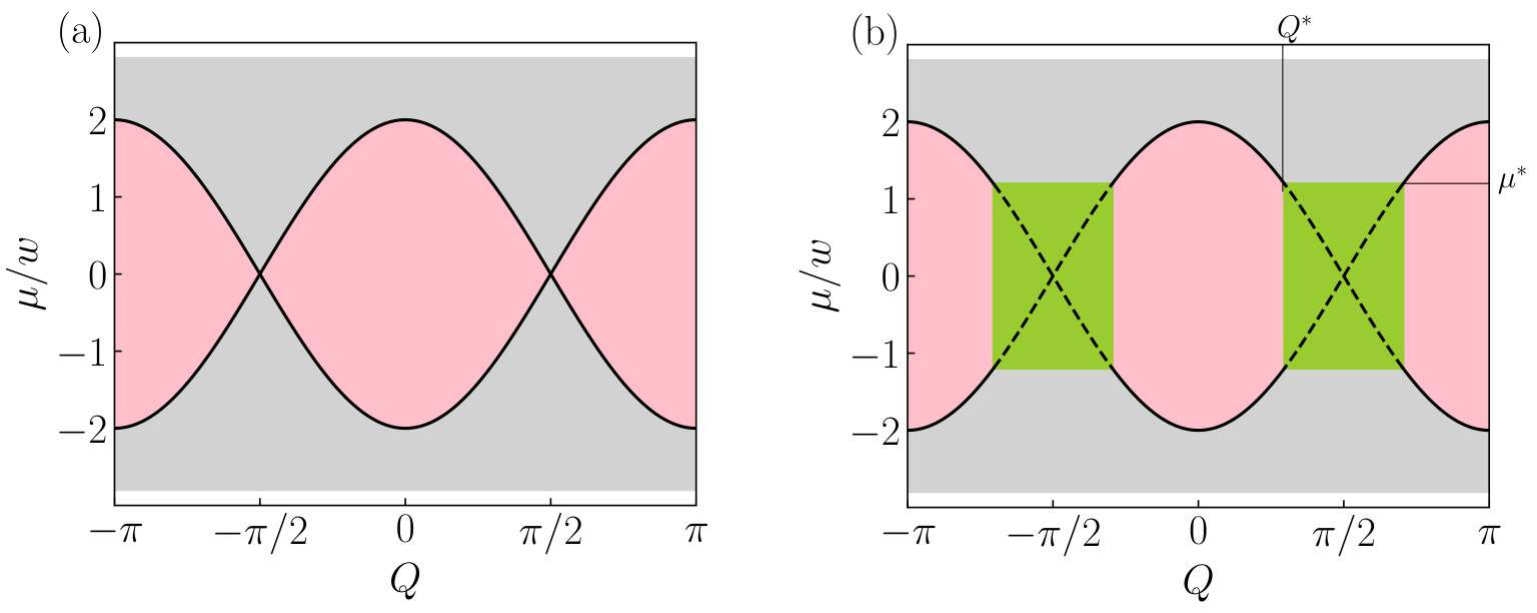}
	\caption{\label{fig_1}Phase diagram of the Kitaev chain as a function of the superconducting phase wavevector~$Q$ and the chemical potential $\mu$, for two regimes of the system (a) $\Delta_{0} = 1.4 w$ and (b) $\Delta_{0} = 0.8w$. The pink and gray   regions correspond to the topological and trivial gapped phases, respectively, while the green regions identify the gapless phases. The   values  $Q^{*}$ and $\mu^*$ characterizing the boundaries of the gapless phases are given in Eqs.(\ref{Q-star}) and (\ref{mu-star}).}
\end{figure}

Gapless phases emerge when $E_{\pm}(k,Q) < 0$ or $E_{\pm}(k,Q) > 0$ for some values of $k$. The conditions for the system to be in a gapless phase are~\cite{FFF2024,FF2024}
\begin{equation}
	\begin{array}{lcccl}
		 && \sqrt{w^{2} - \mu^{2}/4} > \Delta_{0} &\text{and}&w\left|\sin(Q)\right| > \Delta_{0}.
	\end{array}
	\label{gapless_regions}
\end{equation}
In this case, the sectors in $k$ space for paired and unpaired fermions are
\begin{equation}
	S_{p} = \left\{k \, | \, 0 < \left|k\right| < \left|k^{*}_{-}\right| \quad \text{and} \quad \pi - \left|k^{*}_{+}\right| < \left|k\right| < \pi  \right\},
	\label{Sp_sector}
\end{equation}
and
\begin{equation}
	S_{u} = \left\{k \, | \, \left|k^{*}_{-}\right| < \left|k\right| < \pi - \left|k^{*}_{+}\right|\right\}\quad,
\end{equation}
where
\begin{equation}
	k^{*}_{\pm} = \arcsin\left(\frac{\cos(Q)}{\sqrt{1 - \frac{\Delta^{2}_{0}}{w^{2}}}}\right) \pm \arcsin\left( \frac{\mu}{2w\sqrt{1 - \frac{\Delta^{2}_{0}}{w^{2}}}}\right).
	\label{k_boundaries}
\end{equation}
The condition (\ref{gapless_regions})  determines the values of $Q^*$ and $\mu^*$ at which the system is at the boundary between the gapped and gapless phase, as highlighted in Figure \ref{fig_1}{b}. These values are given by
\begin{equation}
	Q^{*} = \arcsin\left(\frac{\Delta_{0}}{w}\right)
	\label{Q-star}
\end{equation}    
and
\begin{equation}
	\mu^{*} = 2 \sqrt{w^2-\Delta_0^2} \quad.
	\label{mu-star} 
\end{equation}
Note that  the system is in the gapless phases if $Q^{*} < \left|Q\right| < \pi - Q^{*}$ and $|\mu|< \mu^*$.

\section{Results}
\label{sec3}
Here, we present our results about the electron   density
\begin{equation}
	\rho = \langle c^{\dagger}_{j}c_{j} \rangle  \quad,
	\label{correlations}
\end{equation}
where $\langle \cdots \rangle$ denotes the expectation value taken with respect to the ground state (\ref{ground_state}). In order to explicitly evaluate $\rho$, one can  re-express $c_{j} = N_{s}^{-1/2}\sum_{k \in BZ}e^{ikj}c_{k}$ in terms of its Fourier modes $c_k$'s, and exploit the correlation functions in $k$-space, where the Brillouin zone sectors $S_{p}, S_{e}$ and $S_{h}$ contribute  differently\cite{FF2024}. In particular, it turns out that only   the sector $S_{p}$ of Cooper pairs  contributes to the density. After taking the  thermodynamic limit,  $\rho$ can be therefore re-expressed as
\begin{equation}
	\rho = \frac{1}{2} + \Delta \rho \quad,
	\label{Normal_C}
\end{equation}
where
\begin{equation}
	\Delta \rho = - \frac{1}{4\pi} \intop_{S_{p}} dk \frac{\xi(k,Q)}{  \sqrt{ \xi^{2}(k,Q)   + \left|\Delta(k)\right|^{2}}     },
	\label{Delta-rho}
\end{equation}
describes the deviation from the half filling density value $1/2$.
We recall that, if the system is in a gapped phase, the $S_p$ sector coincides with the entire Brillouin zone (see Eq.(\ref{sectors-gapped})), while in the gapless phase $S_p$ is strictly smaller than the Brillouin zone and is controlled by the chemical potential $\mu$ and the modulation wavevector $Q$ (see (\ref{Sp_sector})).  \\

\begin{figure}
\centering
\includegraphics[scale = 0.5]{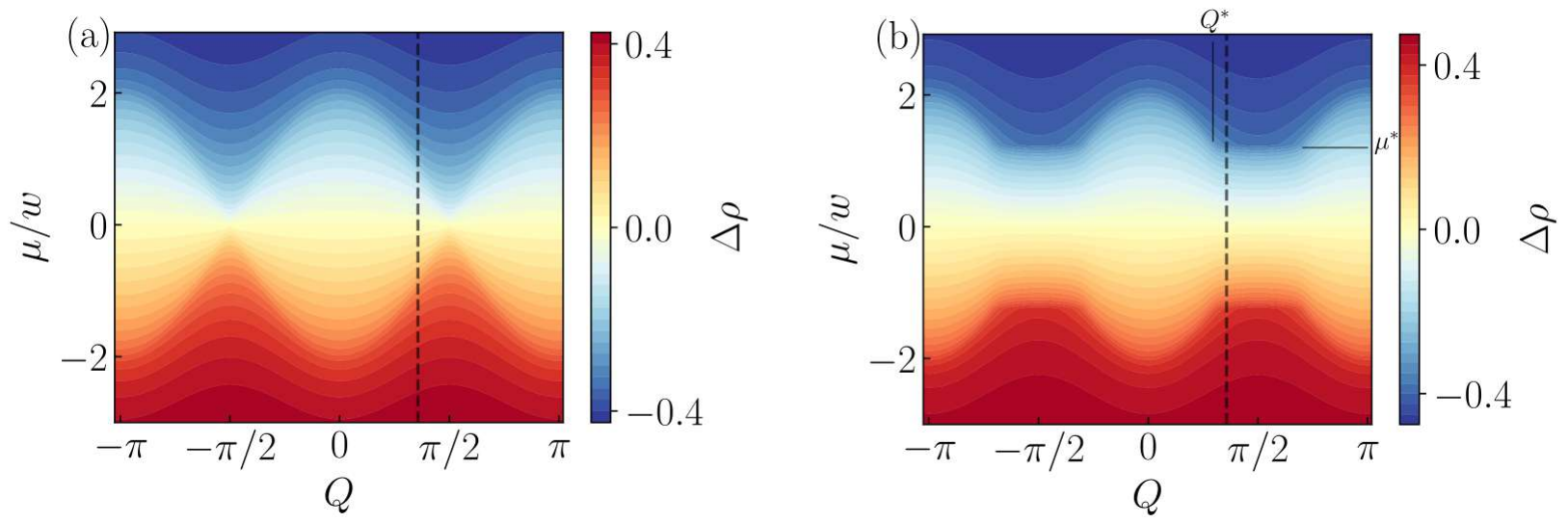}
\caption{\label{fig_2} Contour plot of the density deviation $\Delta \rho$ as a function of $Q$ and  $\mu$. (a) Regime of only gapped phases with $\Delta_0 = 1.4w$. (b) Regime where gapless pheses emerge for some values of $Q$ and $\mu$, with $\Delta_0 = 0.8w$. The values of $Q^{*}$ and $\mu^{*}$ signal the location of the gapless boundaries and are given by (\ref{Q-star}) and (\ref{mu-star}), respectively. The vertical dashed line  corresponds to  a cut at $Q =  \pi/3$ (see Figure \ref{fig_3}).}
\end{figure}

Figure~\ref{fig_2} shows   $\Delta \rho$   as a function of the modulation wavevector $Q$ and the chemical potential~$\mu$. The two panels (a) and (b) refer to the same two values   $\Delta_0 = 1.4w$ and   $\Delta_0 = 0.8w$ used in the two panels of   Figure~\ref{fig_1}, respectively. 
The first feature one can notice from Figure~\ref{fig_2} is that, for $\mu=0$,  one has $\Delta \rho=0$, regardless of the value of the spatial modulation $Q$ and the superconducting pairing strength $\Delta_0$. This means that $\mu=0$ always corresponds to the half filling density value $\rho=1/2$. This property can be shown with a little algebra from Eq.(\ref{Delta-rho}), and originates from the chiral  symmetry of the Hamiltonian (\ref{Ham-Kitaev}), which at $\mu=0$ is invariant  under the (anti-unitary) transformation $
\mathcal{S} c_{j} \mathcal{S}^{-1}  =  (-1)^{j}c^{\dagger}_{j}$~\cite{FF2024}.
The second noteworthy aspect in the behavior of the electron density shown in Figure~\ref{fig_2} is that it straightforwardly reflects the ground state phase diagrams shown in  Figure~\ref{fig_1}. In particular, in Figure~\ref{fig_2}{a}   the density behavior enables one   to clearly distinguish  the separatrix Eq.(\ref{separatrix}) between the two gapped phases (black curves of Figure~\ref{fig_1}{a}), while Figure~\ref{fig_2}{b} allows one to identify the wavevector $Q^{*}$ and the chemical potential value $\mu^*$ (see Eqs.(\ref{Q-star}) and (\ref{mu-star})) characterizing  the onset of the gapless phases shown in Figure~\ref{fig_1}{b}.\\

\begin{figure}
\centering
\includegraphics[scale = 0.68]{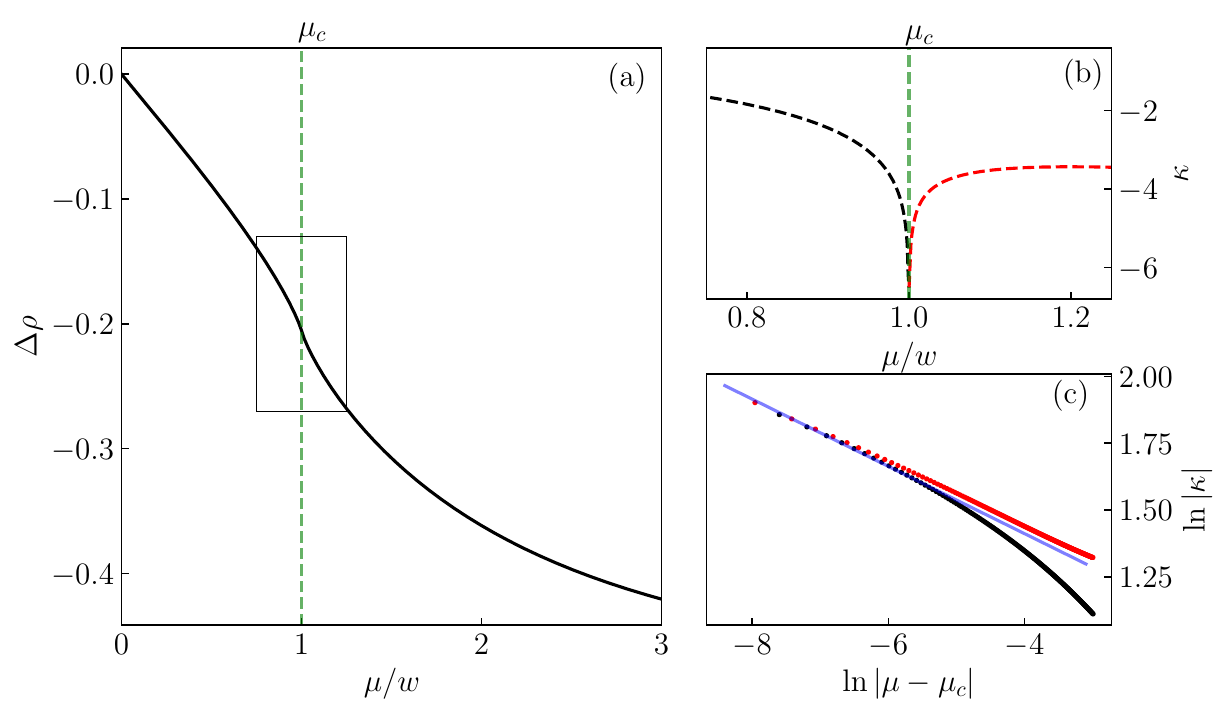}
\caption{\label{fig_3}  Cut of Figure \ref{fig_2}{a} at $Q = \pi/3$. (a) The density deviation $\Delta \rho$ is shown as a function of $\mu$. (b) The compressibility (\ref{kappa-def})  is plotted as a function of $\mu$ in the range within the black rectangle in panel (a). (c) The log-log plot of the compressibility shows a symmetric power law singularity behavior for the compressibility $\kappa \sim |\mu-\mu_c|^b$, where  black and red data  correspond to the ranges $\mu < \mu_{c}$ (topological gapped phase) and $\mu > \mu_{c}$ (trivial gapped phase), respectively.   The ten closer points to $Q_{c}$ were fitted with $\ln |\kappa| = a - b\ln \left|\mu-\mu_{c}\right|$, where $a \sim 0.91$ and $b \sim 0.13$.   }
\end{figure}

In order to gain further information from the  behavior of the density $\Delta \rho$, we have analyzed its $\mu$-dependence  at a fixed value $Q = \pi/3$ of the modulation wavevector, highlighted by a black vertical dashed line in both panels of Figure \ref{fig_2}. We start by examining the case $\Delta_{0} > w$. Figure \ref{fig_3}{a} represents the cut of Figure \ref{fig_2}{a} at $Q=\pi/3$ as a function of $\mu$,  in the range $0 < \mu < 3 w$, and the vertical green dashed line highlights the value of $\mu_c=\mu_c^+(\pi/3)$ (see Eq.(\ref{separatrix})), corresponding to the separatrix between the trivial and the topological gapped phases. A close inspection of the curve shows that it  exhibits a kink at the critical value $\mu_{c}$ . However, such a singular behavior can be better revealed by analyzing the compressibility
\begin{equation}
\kappa=\frac{1}{\rho^2}\frac{\partial \rho}{\partial \mu} \label{kappa-def} \quad,
\end{equation}
which is shown in Figure \ref{fig_3}{b}. A divergence at $\mu_{c}$ is clearly visible in $\kappa$, from both the   topological and the trivial  side of the transition, indicated by the black and red dashed curves in Figure~\ref{fig_3}{b}, respectively. By analyzing such a divergence, we have found   a power-law behavior $\kappa \sim 1/|\mu-\mu_c|^b$ for $\mu\rightarrow \mu_c$, characterized by the {\it same} exponent $b$ from both sides of the transition. This is clearly shown in the log-log plot of Figure~\ref{fig_3}{c}, which illustrates  $\ln |\kappa|$, as a function of $\ln \left|\mu - \mu_{c}\right|$. Here, black dots represent values   for $\mu < \mu_{c}$ and red dots for $\mu > \mu_{c}$, and   they both overlap onto the same   dashed-dotted blue line, which represents a linear fit of the ten points closest to $\mu_{c}$, returning a value $b \simeq 0.13$ for the exponent, for the above model parameters. This symmetrical divergence from both the topological and the trivial side is consistent with previous studies on  correlation functions in topological systems\cite{sigrist_2017}, and highlights the fact that, differently from the conventional Landau phase transition scheme,  two topologically distinct phases are not straightforwardly identified by the onset of an order parameter.\\

\begin{figure}
\centering
\includegraphics[scale = 0.68]{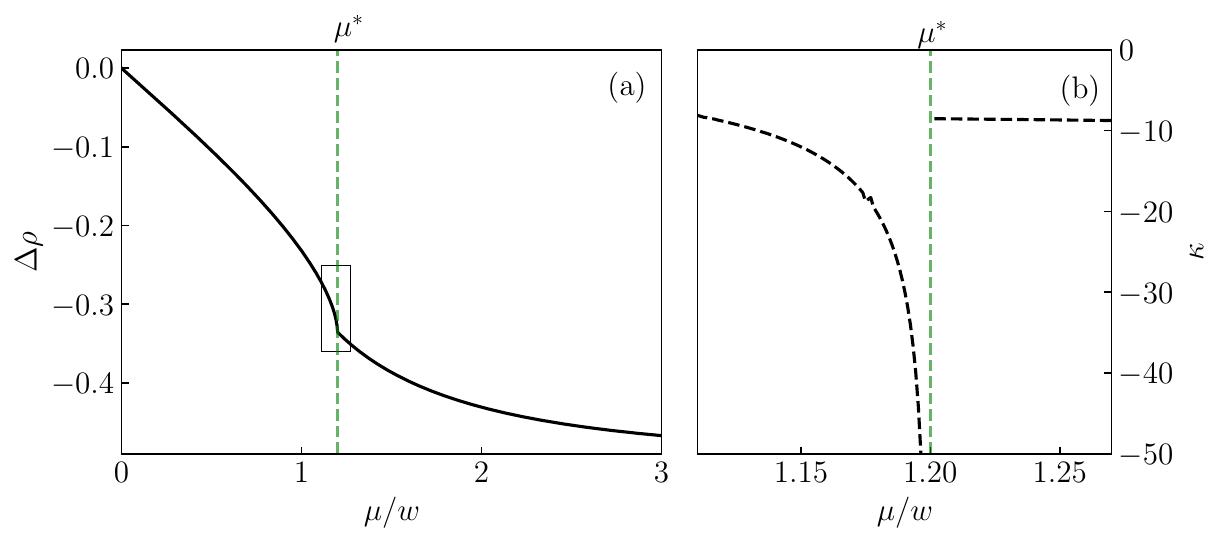}
\caption{\label{fig_4}  Cut of Figure \ref{fig_2}(b) at $Q = \pi/3$. (a) The density deviation $\Delta \rho$ is shown as a function of $\mu$. The vertical green dashed line corresponds to the value $\mu^{*}$ given by (\ref{mu-star}), which is the boundary between gapped and gapless phases. (b) The compressibility (\ref{kappa-def}) is plotted as a function of $\mu$ in the range within the black rectangle in panel (a). An asymmetric jump can be seen at $\mu^*$, namely from the gapped side of the transition ($\mu>\mu_c$) $\kappa$ tends  to a finite value, while it diverges from the gapless side ($\mu<\mu_c$).}
\end{figure}

Let us now turn to the regime $\Delta_0<w$, and analyze the gapped-gapless superconductor transition.
 Figure \ref{fig_4}{a} shows the $Q$-cut of the contour plot in Figure \ref{fig_2}{b} at $Q=\pi/3$. Here, the vertical green dashed line highlights  the boundary $\mu^{*}$ between the   trivial gapped  phase and the gapless   phase, given by Eq.(\ref{mu-star}) and shown also in Figure \ref{fig_2}{b}. As one can see, the kink in the curve Figure \ref{fig_4}{a} is now more pronounced than the one in Figure \ref{fig_3}{a}. The reason is revealed by the analysis of the compressibility (\ref{kappa-def})  displayed in Figure \ref{fig_4}{b}: In striking contrast to the case of the transition across two gapped phases,   the compressibility across the gapped-to-gapless transition exhibits an {\it asymmetric jump}   at $\mu^{*}$. Specifically, while  from the   gapped side of the transition ($\mu>\mu^*$) $\kappa$ is {\it finite},  from the gapless side of the transition ($\mu<\mu^*$) it diverges. This  effect is similar to the one predicted in Ref.\cite{FF2024} for the anomalous correlator, as a function of $Q$ though.

\subsection{Case $Q = \pm \pi/2$}
 We have already observed above that for $\mu=0$ the integral in Eq.(\ref{Delta-rho}) vanishes, reflecting the chiral symmetry acquired by the Hamiltonian, and yielding the half filling value $\rho=1/2$ for the density, at any value of $Q$ and $\Delta_0$. There exists another special parameter value, at which an analytical expression for $\Delta \rho$ can be found, namely   $Q = \pm \pi/2$. Indeed in this case the model (\ref{Ham-Kitaev}) exhibits another symmetry, i.e. it is invariant under the   (unitary) spatial inversion   
$\mathcal{I}c_{j}\mathcal{I}^{-1} =c_{-j}$~\cite{FF2024}. This enables one to rewrite Eq.(\ref{Delta-rho}) as
\begin{equation}
	\Delta \rho =   \frac{\text{sgn}\left(\mu\right)}{\pi} F\left( \alpha ; - 4 \delta_{\mu}  \right) \quad,
	\label{Density_Q}
\end{equation}
where $F(\alpha;- 4\delta_{\mu})$ is the elliptic integral of the first kind.  
Here, $\delta_{\mu} = \Delta^{2}_{0}/\mu^{2}$, while $\alpha$ takes different values depending on the phase. Specifically, for the gapped phase, $\alpha=\pi/2$, while for the gapless phase one has
\begin{equation}
\alpha= \arcsin \left( \frac{ |\mu| }{2\sqrt{w^2-\Delta_0^2}} \right) \quad. \label{k-star-mu-def}  
\end{equation}

\section{Discussion and Conclusions}
\label{sec4}
We have investigated the 1D Kitaev chain model in the presence of a phase modulation of the superconducting pairing, which describes a current flowing through the chain. Such modulation is known to induce two types of effects. On the one side, it  affects the customary band topology transition by modifying the separatrix between the topological and trivial gapped phases, where the spectral gap closes directly. On the other hand, in the physically realistic parameter regime $\Delta_0<w$, it can lead to a Lifshitz transition, i.e. a topological change in the Fermi surface. This corresponds to   the appearance of a gapless superconductor phase  (see Figure \ref{fig_1}), characterized by an indirect spectral gap closing.   \\

Here, we have analyzed the behavior of the electron density $\rho$ and the compressibility $\kappa$ across these two types of topological transitions. 
We have found that the behavior of the density   as a function of the phase modulation wavevector $Q$ and the chemical potential~$\mu$ (Figure \ref{fig_2}) is quite informative  to infer the phase diagram of Figure \ref{fig_1}. In particular, in the regime $\Delta_0>w$, it enables one to identify the separatrix (\ref{separatrix}) between the gapped phases (Figure \ref{fig_1}{a}), while in the regime $\Delta_0<w$, one can extract the boundary values  $Q^*$ and $\mu^*$ (see Eqs.(\ref{Q-star}) and (\ref{mu-star})) characterizing the gapless phase. Moreover, the analysis of the compressibility (\ref{kappa-def}) enables one to operatively distinguish between the two types of transitions. Indeed,  the customary band topology transition between topologically different gapped  phases is characterized by a {\it symmetric} power law divergence of $\kappa$ at the critical value $\mu_c$  (see Figure \ref{fig_3}{b}), with the same  exponent both from the trivial and the topological side of the transition.  In contrast, across the gapped-gapless transition, $\kappa$ is characterized by an {\it asymmetric jump} at $\mu^*$: While it tends to a finite value from the gapped side, it diverges from the gapless side of the transition (see Figure \ref{fig_4}{b}).\\

We conclude by outlining possible experimental setups, where our results could be tested.  
Two types of platforms seem  promising for the realization of 1D topological superconductors. The first one is based on InSb and InAs  nanowires   proximized by a  superconducting film (e.g., Al or Nb) and exposed to a  magnetic field~\cite{kouwenhoven2012,furdyna2012,heiblum2012,kouwenhoven2018,yacoby2014,yu2021non}, while the second one consists of ferromagnetic atom chains deposited on a superconducting film~\cite{yazdani2014,tang2016,loss-meyer_2016}. 
Electron density profile in semiconductors can be locally probed with various techniques such as  atomic force microscopy, capacitance voltage measurements and  x-ray reflectivity\cite{chaudhari_2000,garnett_2009,marisol_2013,wielgoszewski_2015,park_2024}. The chemical potential can be controlled by a gate voltage, and the compressibility of electron gases in semiconductors can be measured with capacitive techniques\cite{eisenstein_1994}.
Moreover, scanning tunneling microscopy has been proposed as a technique to measure local  correlation functions in magnetic atom  chains~\cite{choy2011,nadjperge2013}.


\acknowledgments
F.G.M.C. acknowledges financial support from ICSC Centro Nazionale di Ricerca in High-Performance Computing, Big Data, and Quantum Computing (Spoke 7), Grant No. CN00000013, funded by European Union NextGeneration EU. \,\, F.D. acknowledges financial support  from the TOPMASQ project, CUP E13C24001560001, funded by the  National Quantum Science and Technology Institute (Spoke 5), Grant No. PE0000023, funded by the European Union – NextGeneration EU.

\bibliography{Biblio}

\end{document}